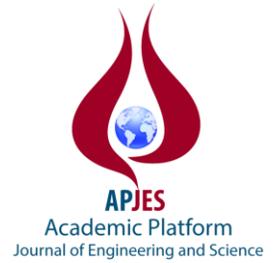

# Scheduling Cutting Process for Large Paper Rolls

* [1]Mehmet E. Aydin and [2] Osman Taylan

[1]Dept. of Computer Science and Technology, University of Bedfordshire, Luton, UK
*[2]Dept. of Industrial Engineering, King Abdulaziz University, Saudi Arabia

*Corresponding author: Address: Dept. of Industrial Engineering, King Abdulaziz University, Saudi Arabia
E-mail address: mehmet.aydin@beds.ac.uk

## Abstract

Paper cutting is a simple process of slicing large rolls of paper, jumbo-reels, into various sub-rolls with variable widths based on demands risen by customers. Since the variability is high due to collected various orders into a pool, the process turns to be production scheduling problem, which requires optimisation so as to minimise the final remaining amount of paper wasted. The problem holds characteristics similar one-dimensional bin-packing problem to some extends and differs with some respects. This paper introduces a modelling attempt as a scheduling problem with an integer programming approach for optimisation purposes. Then, a constructive heuristic algorithm revising one of well-known approaches, called Best-fit algorithm, is introduced to solve the problem. The illustrative examples provided shows the near optimum solution provided with very low complexity.

## 1. Introduction

Paper machines produce large reels of paper, with a fixed width which is called deckle; the reels are later cut into rolls of paper sized to customer specifications where each customer order defines the quantity, product type, the roll width and diameter. A typical example is the cutting of a wide paper reel (jumbo-reel) into smaller paper rolls, which are either end-customer rolls or intermediate products waiting for further processing, such as printing, coating or cutting. In paper industry; the trim-losses problem appears when customer's demand is to be satisfied in a paper converting mill, where a set of product paper reels need to be cut from raw-paper reels. In the simplest terms, the problem as is called, is to determine the number of logs required and the way each is to be processed in order to satisfy a set of customer orders in an economical manner. The main objective in such problems is to minimize the trim losses, residuals, while producing the rolls according to customer order specifications.

The problem is modelled in integer programming form and therefore gives rise to a difficult combinatorial nature. The one-dimensional cutting stock problem (CSP) or trim problem is introduced in Operations Research for the purpose of modelling roll-cutting problem. Although the formulation has first been brought in by Nobel Prize winner Kantorovich in 1939 in linear programming form [9], as a combinatorial optimisation model, one-dimensional bin-packing problem can be used to better implement this particular problem. In order to ease the practicality, the problem can be considered as a production scheduling problem in which all demand risen by the customers to be delivered with a minimum trim-loss and maximum satisfaction. An intermediatory cutting process is designed to cut the jumbo-reels into rolls which come up with non-standard widths, which are determined based on the size of orders, originally done in weights not widths. Figure 1 depicts the main functionality of cutting process in a way that a jumbo-reel with width of *W* is to be cut into *N* sub-rolls with widths varying between $w_1$ to $w_m$ and the residual of *t*. A particular procedure is used to calculate the number of rolls ordered in a non-standard width. References [1], [2] and [3] introduce such procedures how to convert weight-based orders into width-based ones. Both [1] and [2] also provide details of optimisation with simulated annealing. However, none of above mentioned papers discusses and introduces any model for scheduling aspect of the cutting process of the jumbo-reels. There is a literature produced in the past as reported in [7], but, not easily accessible.

The rest if this paper is organised as follows: Section 2 introduces an integer programming model in which the schedule how and when to cut the rolls while Section 3 presents a heuristic/greedy constructive algorithm to solve the problem. An illustrative example is exposed with discussions in Section 4 and paper is concluded in Section 5

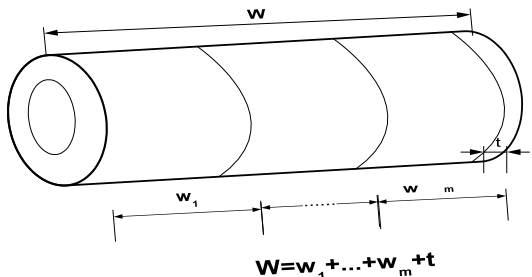

**Fig. 1** A typical jumbo-reel to be cut into rolls and the possible trim loss

## 2. Modelling the schedule of paper-cut

Let *N* be the number of orders collected in the demand pool, where each order is recorded the number of rolls, $R_i$, each with width of $w_i$. The total width to be cut is $W^d$ defined as:

$$W^d = \sum_{i=1}^{N} R_i w_i \quad (1)$$

Once $W^d$ is calculated, then the number of jumbo-reels to be cut, $R^d$, can be determined with:

$$R^d \geq \left\lceil \frac{W^d}{W} \right\rceil \quad (2)$$

where $\lceil . \rceil$ denotes a ceiling function. The objective function is:

$$Z = \min\left( W^d - \sum_{i=1}^{R^d} \sum_{j=1}^{N} w_j r_{ij} y_{ij} \right) \quad (3)$$

*Z* is the overall residual, which ends up as wasted amount to be minimised. The main aim is to develop a cutting schedule which makes sure that the total amount of demand is met; therefore the following equation needs to be satisfied.

$$\sum_{j=1}^{N} w_j r_{ij} y_{ij} \leq W \quad \forall i \in R^d \quad (4)$$

where *N* is the number of orders, $R^d$ is also the number of slots per cutting schedule, $w_j$ is the width of $j^{th}$ order, $r_{ij}$ is the number of rolls cut for $j^{th}$ order within $i^{th}$ slot while $y_{ij}$ is a binary variable to identify whether any cut is made for $j^{th}$ order in $i^{th}$ slot. This constraint makes sure the total width cut per slot cannot be greater than the gross-width, *W*.

Each slot of cutting schedule cannot be a non-cutting slot, therefore:

$$\sum_{j=1}^{N} y_{ij} \geq 1 \quad \forall i \in R^d \quad (5)$$

where $y_{ij} \in [0, 1]$.

As the cut of any particular order can be delivered within multiple slots, every order must be completed in size by the end of the schedule. Therefore, the sum of all completed rolls is required to be equal to the original size of the order, $R_i$.

$$\sum_{j=1}^{R^d} r_{ij} y_{ij} = R_i \qquad \forall i \in N \quad (6)$$

Reviewing the whole model, it is observed that there are two decision variables; one is $y_{ij}$, fully independent and the other is $r_{ij}$, a semi-independent variable, which is fine-tuned depending on the level of $y_{ij}$. Since the problem is considered as a complete cutting schedule, both of the decision variables turn in form of matrixes; $Y$ and $R$.

Therefore, the problem is defined as a function of $Y$ and $R$ resulting in real numbers in the following form:

$$f:(Y,R) \longrightarrow \Re \quad (7)$$

where $Y$ is matrix of binary decision variables used to identify the slots of the schedule dedicated to the orders, while $R$ is a matrix of integers to count the numbers of the rolls to be cut at each slot. Therefore, the main purpose of the model turns to make a decision on $Y$ and $R$ to measure the performance of the state of the model.

This problem can be considered as an implementation of one-dimensional bin-packing problem in which the number of bins packed in is minimised. The width of the jumbo-reel can be considered as the capacity of a bin while each of order will be a group of items to be packed in the bins. The main difference would be the objective function, which is to minimise the wasted amount in this problem, Eq:(3), while it is to minimise the number of bins in bin-packing problem, the equality form of Eq:(2). Therefore, the constituting complexity of paper-cutting problem is NP-Hard as is in one-dimensional bin-packing problem.

## 3. A heuristic approach for solving the problem

There are various heuristic approaches to solve bin-packing problems alongside global optimisation methods, where some of them are constructive algorithms and some are explorative. Among the constructive ones, First-fit and Best-fit algorithms [4] [5] are quite commonly used ones. On the other hand, metaheuristic approaches are implemented to solve bin-packing problems similar to other NP-Hard combinatorial optimisation problems [6].

The approach in this section proposes another constructive algorithm, which works similar to Best-fit algorithm in a way that the least capacity can hold an item is looked for through the algorithm. However, it does not do that for all orders, but the half of the orders. The approach focuses on domain-specific information in which the problem can be easily and straightforwardly solved. It requires dividing the whole pool of demands/orders, **P** into two sub-pools; one is denoted with $P_w$ to hold the orders wider than the half of the width of jumbo-reel and the other is denoted with $P_n$ to keep those orders narrower than the half-width. Consequently, a particular order is classified based on the width required, whether or not it is wider than the half-width of the jumbo-reel. The wider ones are classified into $P_w$, while the rest are grouped in $P_n$. The members of $P_w$ cannot fit in a jumbo-reel more than once, but the members of $P_n$ can fit in.

The algorithm has two main steps; classification step and coupling step. It, first, starts with classifying the orders, and then moves to coupling stage, where the best fitting couples from both sub-pools are looked for. Once coupling is completed, then jumbo-reels are assigned. The following steps are to indicate the procedure of the algorithm roughly.

1. Pick up the order with widest width in $P_w$,
2. Couple it with the best fitting member(s) of $P_n$ so as to get minimum waste,
3. Repeat the same action if more can be coupled in the same scheme,
4. Remove all coupled from both pools,
5. Repeat this action until one of the pools dried out,

6. If the dried out pool is $P_w$ then repeat the items 1-5 for $P_n$ until all members are removed completely,
7. Assign a jumbo-reel to each complete couple,
8. If $P_n$ dried out before $P_w$,
9. Calculate the total waste with Eq:(3).

There is a similarly operating algorithm reported in literature [8] applied to two- dimensional strip backing problem, which differs from the proposed one in a way that it applies First-fit algorithm after splitting the whole pool, but, the proposed algorithm does not apply First-fit at all. The time complexity of this algorithm is, normally, the half of the Best-fit algorithm since it does not look for all orders, but, for those classified in $P_n$ pool. It picks up items from $P_w$ and looking for best matches within $P_n$, where the complexity remains depending on the size of $P_n$, which is decreasing by removing the matched items. Therefore, the complexity will be O(s log s), where s = | $P_n$ | returning the size of content $P_n$.

## 4. Illustrative Examples

In this section, two illustrative examples are presented to show the model developed is well-working and ready to be used for further actions in optimization. The first example, the first 10 orders taken from [1], is based on the data tabulated in Table 1, which consists of 10 separate orders, each includes the width, the weight and the number of rolls required. The orders are collected in the pool based on planning policy and capacity, whose the size of pool can change based upon. The width of a jumbo- reel is known as 201.0 cm. 1 cm is ignored since it would be wasted through cutting operations. The second example presented in Table 2 comprising 18 orders is a rather based on real data, where the width of jumbo-reel is known as 2500mm.

**Table 1** A list of 10 orders pooled

| Order ID | Width (cm) | Weight (kg) | R |
|---|---|---|---|
| $D_1$ | 55.0 | 2035 | 6 |
| $D_2$ | 145.0 | 5365 | 6 |
| $D_3$ | 50.0 | 2267 | 8 |
| $D_4$ | 150.0 | 1125 | 2 |
| $D_5$ | 135.0 | 5108 | 6 |
| $D_6$ | 80.0 | 5386 | 12 |
| $D_7$ | 105.0 | 4030 | 6 |
| $D_8$ | 90.0 | 2842 | 5 |
| $D_9$ | 100.0 | 3158 | 5 |
| $D_{10}$ | 55.0 | 8137 | 24 |

**Table 2** A list of 18 orders pooled

| Order ID | Width (cm) | R | Order ID | Width (cm) | R |
|---|---|---|---|---|---|
| $D_1$ | 1470 | 7 | $D_{10}$ | 11120 | 21 |
| $D_2$ | 1030 | 20 | $D_{11}$ | 1150 | 9 |
| $D_3$ | 1450 | 24 | $D_{12}$ | 1350 | 9 |
| $D_4$ | 1050 | 12 | $D_{13}$ | 1330 | 14 |
| $D_5$ | 1080 | 11 | $D_{14}$ | 1180 | 9 |
| $D_6$ | 1410 | 11 | $D_{15}$ | 1300 | 9 |
| $D_7$ | 1400 | 12 | $D_{16}$ | 1250 | 27 |
| $D_8$ | 1100 | 11 | $D_{17}$ | 950 | 17 |
| $D_9$ | 1370 | 7 | $D_{18}$ | 1550 | 17 |

Applying the heuristic algorithm, the members of $P_w$ are identified as $D_2$, $D_4$, $D_5$ and $D_7$, while the demands classified in $P_n$ are $D_1$, $D_3$, $D_6$, $D_8$, $D_9$ and $D_{10}$. The solution produced with the heuristic algorithm is given in Table 4 with an overall wasted amount of 230 cm out of 34 jumbo-reels cut. Obviously, the theoretical number of jumbo-reels calculated with Eq. (1) is 32 with a residual of 160 cm. This can be considered as the lower boundary. The theoretical residual can be saved, while the found residual cannot be. On the other hand, First-fit algorithm produces a solution of 36 jumbo-reels with a residual of 630 cm, where 100 cm still can be used for further cuts, but, 530 cm will most likely be wasted.

**Table 3** A possible schedule for cutting rolls for orders tabulated in Table(1)

| Paired Orders | Rolls Required | Waste per Roll | Total Waste |
|---|---|---|---|
| $D_4$ (2) + $D_3$ (2) | 2 | 0 | 0 |
| $D_2$ (6) + $D_1$ (6) | 6 | 0 | 0 |
| $D_5$ (6) + $D_{10}$ (6) | 6 | 10 | 60 |
| $D_7$ (5) + $D_8$ (5) | 5 | 5 | 25 |
| $D_7$ (1) + $D_6$ (1) | 1 | 15 | 15 |
| $D_9$ (2) + $D_9$ (2) | 2 | 0 | 0 |
| $D_9$ (1) + $D_3$ (2) | 1 | 0 | 0 |
| $D_6$ (9) + $D_{10}$ (9) + $D_{10}$ (9) | 9 | 10 | 90 |
| $D_6$ (2) + $D_3$ (2) + $D_3$ (2) | 2 | 20 | 40 |

The second example is solved with the proposed heuristic algorithm as well as first-fit and linear programming. Applying proposed heuristic divides the pool into $P_w$ and $P_n$ as described above, where $P_w$ consists of $D_{18}$, $D_1$, $D_3$, $D_6$, $D_7$, $D_9$, $D_{12}$, $D_{13}$, $D_{15}$ and $D_{16}$ while $P_n$ comprises $D_{14}$, $D_{11}$, $D_{10}$, $D_8$, $D_5$, $D_4$, $D_2$ and $D_{17}$. The whole solution is presented in Table 4 with a performance of 124 jumbo-reels used and trim-loss of 2620mm, where 1250mm is available not to be wasted and used for further orders. Apparently, this is the optimum solution found with linear programming while First-fit found 128 jumbo-reels to be used with 3 of them to be cut in half. This demonstrates the efficiency both the model and the algorithm.

**Table 4** A possible schedule for cutting rolls for orders tabulated in Table(2)

| Paired Orders | Rolls Required | Waste per Roll | Total Waste |
|---|---|---|---|
| $D_{18}$ (17) + $D_{17}$ (17) | 17 | 0 | 0 |
| $D_1$ (7) + $D_2$ (7) | 17 | 0 | 0 |
| $D_3$ (12) + $D_4$ (12) | 12 | 0 | 0 |
| $D_3$ (12) + $D_2$ (12) | 12 | 20 | 240 |
| $D_6$ (11) + $D_5$ (11) | 11 | 10 | 110 |
| $D_7$ (11) + $D_8$ (11) | 11 | 0 | 0 |
| $D_7$ (1) + $D_2$ (1) | 1 | 70 | 70 |
| $D_9$ (7) + $D_{10}$ (7) | 7 | 10 | 70 |
| $D_{12}$ (9) + $D_{11}$ (9) | 9 | 0 | 0 |
| $D_{15}$ (9) + $D_{14}$ (9) | 14 | 50 | 700 |
| $D_{13}$ (14) + $D_{10}$ (14) | 9 | 20 | 180 |
| $D_{16}$ (13) + $D_{16}$ (13) | 13 | 0 | 0 |
| $D_{16}$ (1) | 1 | 1250 | 1250 |

## 5. Conclusion

Paper cutting is a process of slicing large rolls of paper, jumbo-reels, into various sub-rolls with variable widths based on demands raised. Developing a cutting schedule turns to be a very complicated optimisation problem, which needs serious attempts to solve it. As the problem is in NP-hard nature, larger size of the problem will not plausibly solved with global optimisation approaches. This paper introduces a modelling attempt for scheduling the paper cutting process with integer programming approach for optimisation purposes. A heuristic algorithm is also introduced with a reasonable performance. The problem is going to be attempted to solve with other metaheuristic approaches in the futu